\documentclass[runningheads]{llncs}
\usepackage{graphicx}
\usepackage[utf8x]{inputenc}
\usepackage[ampersand]{easylist}
\usepackage{graphicx}
\begin{document}
\title{Conversation-Based Complex Event Management in Smart-Spaces}
%\title{Exploring Complex Event Management
%in Smart-Spaces through a
%Conversation-Based Approach}
%
\titlerunning{Conversation-Based Complex Event Management in Smart-Spaces}
\author{André Sousa Lago\inst{1}\orcidID{0000-0002-4534-9180} \and
Hugo Sereno Ferreira\inst{2}\orcidID{0000-0002-4963-3525}}
\authorrunning{André S. Lago, Hugo S. Ferreira}
\institute{Faculty of Engineering of University of Porto, 4200-465 Porto, Portugal \and
Department of Informatics Engineering, Faculty of Engineering of University of Porto, 4200-465 Porto, Portugal}
\maketitle              % typeset the header of the contribution
\begin{abstract}
    Smart space management can be done in many ways. On one hand, there are conversational assistants such as the Google Assistant or Amazon Alexa that enable users to comfortably interact with smart spaces with only their voice, but these have limited functionality and are usually limited to simple commands. On the other hand, there are visual interfaces such as IBM's Node-RED that enable complex features and dependencies between different devices. However, these are limited since they require users to have a technical knowledge of how the smart devices work and the system's interface is more complicated and harder to use since they require a computer.
    This project proposes a new conversational assistant - Jarvis - that combines the ease of use of current assistants with the operational complexity of the visual platforms. The goal of Jarvis is to make it easier to manage smart spaces by providing intuitive commands and useful features. Jarvis integrates with already existing user interfaces such as the Google Assistant, Slack or Facebook Messenger, making it very easy to integrate with existing systems. Jarvis also provides an innovative feature - causality queries - that enable users to ask it why something happened. For example, a user can ask "\textit{why did the light turn on?}" to understand how the system works.

\end{abstract}
\section{Introduction}\label{sec:introduction}

\subsection{Internet of Things}

The Internet of Things, or IoT, is the networked connection of everyday objects, which is often equipped with a collective sense of intelligence~\cite{Xia2012}. The integration of such objects creates a huge range of distributed systems that are able to interact with the environment and human beings around them in a lot of different ways.

The flexibility of IoT systems has enabled their use across many different product areas and markets, including smart homes, smart cities, healthcare, transportation, retail, wearables, agriculture and industry~\cite{Rahul2017}.

One common application of IoT are \textbf{customized smart spaces} as they make it possible for almost everyone to create a customized IoT experience based on products and hardware available. By using devices such as RaspberryPi computers or Arduino components, it is possible to even remotely interact with lights, fridges or other devices to perform automated tasks such as turning on lights at a certain time.

The management of an IoT system is very important as it allows users to understand and modify the way the system operates. Besides the raw solution of manually programming the devices to work as intended, there are two possible approaches: visual programming platforms and conversational assistants. These are described thoroughly in the following sections.

\subsection{Visual Programming Paltforms}

Visual programing platforms, or VPPs, are tools that are usually installed in supervisor devices in IoT systems so that they can access all the devices and components in such systems. Because of that, these platforms can offer users with extensive and friendly UIs through which the user can visualize the state of the system, customize its behavior or even configure the system’s devices themselves. Some VPPs even offer integrations with third parties such as Twitter or Instagram, so that their APIs can be used as part of the system’s behavioral rules.

As an example, we will take a look at one of the most popular VPPs (Node-RED), but it is important to note that there are other relevant alternatives, such as the Home Assistant\footnote{\url{https://home-assistant.io/}}.

\textbf{Node-RED}\footnote{\url{https://nodered.org/}} is a tool developed by IBM’s Emerging Technology Services team in 2013, and is now a part of the JS Foundation. Node-RED follows a flow-based programming approach to the management of an IoT system, providing its users with a Node.js-based web application through which they can create, edit and delete system rules and connections in an interface that displays rules and connections as a flow of information, events or actions.

In Node-RED, system devices are represented as colourful nodes that possess specific properties and actions which can be interconnected with other nodes. Similarly, conditions, actions and events are also represented as nodes, which makes it easy to understand how elements can be connected in the platform. Being based in Node.js, all that is required to use Node-RED is to have it running in a supervisor device in the system, such as a Raspberry Pi, and then the platform can be accessed through any regular web browser.

\begin{figure}
    \begin{center}
        \includegraphics[width=0.5\textwidth]{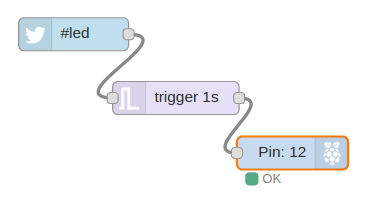}
        \caption{Simple setup of a Node-RED managed system.} \label{fig:nodered-simple}
    \end{center}
\end{figure}

Figure~\ref{fig:nodered-simple} represents a simple application of Node-RED to manage an IoT system. In the example, the user has defined a flow where built-in Node-RED nodes are connected in a flow of information. With blocks similar to those displayed by the figure it is possible to not only send commands to IoT devices but also act upon events received from them, which brings the possibility to create rather complex systems with many dependency rules.

When multiple rules such as the one displayed in Figure~\ref{fig:nodered-simple} are setup in Node-RED, it becomes possible to create complex systems that turn on lights at a certain time, activate devices such as microwave ovens when it is time for breakfast or close the windows if it starts raining.

VPPs can be extremely useful for users of custom smart spaces due to the flexibility and power they provide. However, they possess several disadvantages that make them somewhat hard to use, especially for users that are not entirely comfortable with understanding how certain technologies work.

Let's imagine a Node-RED system, embedded in a user’s home, with multiple devices. Even if there are only a couple of dozens of rules defined, it can be extremely difficult to understand why a certain event took place due to the overwhelming flow that results from them. A single dozen of rules can already result in a Node-RED page that you need to scroll to completely visualize, let alone a couple dozen! The more rules you add to the system, the harder it becomes to look at Node-RED’s interface and understand what the system is able to do, in part because it is impossible to make any other form of  “query” to the platform besides looking at it.

Another major disadvantage of this sort of platforms is that they require the user to sit in front of a computer to set up the system, even if it is for simple tasks. For example, if a user is sitting in his couch, far away from his computer, and thinks that it would be great to have his light turn on as soon as it gets dark outside, he would need to actually get up and go to the computer when there can possibly be a lot of IoT devices around him that he could interact with. Again, this can make these platforms hard or boring to use as it may require a lot of time to perform simple tasks such as the one described.

\subsection{Conversational Assistants}

As an alternative to VPPs there are many conversational assistants, such as the Google Assistant\footnote{\url{https://assistant.google.com/}} or Amazon Alexa\footnote{\url{https://developer.amazon.com/alexa}}, that are capable of answering questions based on knowledge bases, read out emails or notifications and, most importantly, interact with smart spaces. As an example, we will take a deeper look into the Google Assistant as it represents the relevant features that are also present in its alternative products.

There are plenty of ways users can interact with the Google Assistant: the standalone Google apps, built-in integrations with Android (6.0+) and Chrome OS, or with standalone hardware, such as the Google Home. Through these interfaces, it is possible to ask the Assistant questions, ask it to do things and interact with other associated products.

With this, comes one of the interesting use cases of the Assistant - interacting with smart spaces. With the Assistant it is possible to talk to smart space devices such as NEST thermostats\footnote{\url{https://nest.com/thermostats/nest-learning-thermostat/overview/}}, as well as third-party services like Philips Hue\footnote{\url{https://www2.meethue.com/en-us}} bulbs or IFTTT\footnote{\url{https://ifttt.com/}} queue systems, among others\footnote{\url{https://support.google.com/googlehome/table/7401014}}.

The problem with the Assistant’s approach is that these interactions are quite simple, and are mostly directly associated with direct commands or queries to the smart devices. All the Assistant can do with these devices is perform queries like “\textit{is the baby monitor on?}” or “\textit{what is the temperature in the living room?}”, or execute direct actions such as “\textit{turn on the coffee machine}”\footnote{\url{https://store.google.com/us/product/google\_home\_learn?hl=en-US}}.

What this means is that although intelligent assistants like the Google Assistant, Siri and other can make it much more comfortable to interact with smart spaces because they remove the need of touching a physical device, they do not allow you to define rules for how the spaces operate. Saying things like “\textit{everyday at 3pm close the windows}” or  “\textit{when it is raining turn on the porch light}” won’t work on these assistants unless you manually define every single one of them.

Overall, it is easy to understand that although current smart assistants can be very helpful and comfortable to use, they don’t yet have the complexity and completeness that other IoT management systems like Node-RED possess. Also, some of them are limited to a specific range of vendor devices, so there is always a limitation to the customization they can offer.

% \subsection{IoT Communication Protocols}

% When it comes to IoT, there are many protocol definitions that are essential for systems to work, and these can fit into the many layers they are made of. These protocols can be dedicated for IoT, but they can also be general application protocols that happen to be usable in these systems.For example, IoT systems may use the IPv6 protocol for the infrastructure layer, Bluetooth in the communication layer or MQTT for data protocols.

% Mozilla is currently developing a Web Thing API\footnote{\url{https://iot.mozilla.org/wot/}} which is expected to be submitted for approval to the W3C\footnote{\url{https://www.w3.org/}}, a community that brings together companies and experts to develop Web standards. The goal of Mozilla’s Web Thing API is to describe a data model and API that could be used in the context of IoT describe physical devices in a JSON format. This API could then be used together with the communication platforms like MQTT and RabbitMQ to create a more standardized way of establishing communication structures and protocols for IoT systems.

% While the document is still being prepared for submission, it’s principles for communication structures and device identification are already interesting and useful for developing projects. The description of some concepts described by Mozilla’s API can be found below. The JSON code described by the API can fully describe the status and abilities of a “web thing”, which can be useful for example for knowing what sort of actions a certain device is capable of or to read its current status.

\section{Problem Statement}

This section identifies the problem with the current approaches to smart space management, along with the solution proposed by this article.

\subsection{Current Issues}

As presented in the sections above, the current solutions available in the market offer great alternatives for the management of smart spaces, but none of them seems complete as a whole. This is because none of the presented tools simultaneously has these features:

\begin{easylist}[itemize]
    & \textbf{Complex Management:} the ability to perform a wide range of tasks, including direct actions, delayed actions, conditional actions or device interoperability.
    & \textbf{Comfort and ease of use:} the possibility to manage the IoT system with the minnimum possible effort. The maximum comfort would be for the user not to have to move or touch a device in order to get his tasks done, as can happen with voice assistants.
    & \textbf{Understanding system's functioning:} the ability to understand how the configure system works or why something was done. For example, with Node-RED, this is only possible by looking at all the configured rules to figure out which one could have caused somethign to happen. Ideally, all that should be needed is to ask the system why something happen and it should do that search for the user.
\end{easylist}

\subsection{Proposal}

The goal of this project is to develop a conversational bot dedicated to the management of smart spaces that is capable of defining and managing complex system rules, called \textbf{Jarvis}.

Jarvis's abilities reach across different levels of operational complexity, ranging from direct one-time actions (e.g. \textit{"turn on the light}) to repeating conditional actions (e.g. \textit{"when it is raining, close the windows"}). Besides that, Jarvis also lets the user easily change or understand the status of the system, through queries like \textit{"why did the toaster turn on?"}. In that latter case, Jarvis should also possess conversational awareness that allows for chained commands. This particular feature is demonstrated by the following dialogue:

\indent\indent \textbf{User}: “\textit{Why did the toaster turn on?}”

\indent\indent \textbf{Jarvis}: "\textit{You told me to turn it on at 8 AM.}”

\indent\indent \textbf{User}: “\textit{Okay, change it to 7:50 AM.}”

\indent\indent \textbf{Jarvis}: “\textit{Sure, toaster timer was changed.}”

In the example above, the second query of the user wouldn't make sense on its own, however it does make sense as a follow-up to the previous interactions. This can not only be extremely user but also facilitate the user's experience since it avoids repeating information that was already mentioned in the conversation.

To make the bot easy to integrate with current systems, its interface was made through existing platforms like the Google Assistant, Amazon Alexa, Facebook Messenger, Slack, among others. This range of integrations give the bot the ability to interact with users via voice or text queries. The project's code should also be available under an open GitHub repository\footnote{\url{https://github.com/andrelago13/jarvis}}.

\subsection{Research Questions}

With the development and documentation of this project, we aim to answer the following research questions:

\begin{enumerate}
    \item \textbf{"How can direct, delayed, period and repeating actions be implemented?":} this type of features can be very helpful in a voice assistant and make it a very powerful tool for users. With the support for these features, queries such as "\textit{turn on the light}", "\textit{turn on the light in 5 minutes}" or "\textit{turn on the light everyday from 10am to 11am}" become possible.

    \item \textbf{"Can we use contextual awareness for IoT management?":} contextual awareness can help make conversations feel more natural and intuitive as they are closer to human interactions and prevent the user from having to repeat certain commands or phrases. 
    
    \item \textbf{"Can a conversational assistant handle event management?":} events make it possible for certain devices to have a behavior that depends on other devices without being directly connected to them. Because of that, being able to support this sort of features can make a system truly powerful and useful. An example of an event query would be "\textit{Turn on the kitchen light if the kitchen motion sensor is activated}".
    
    \item \textbf{"How can a conversational bot solve causality queries?":} a query such as "\textit{how did that happen?}" can be very helpful for a user to understand how a system is operating or to change the rules previously created. As will be seen, this is not a trivial problem since it has many possible approaches.
\end{enumerate}

\section{Developed Solution}

This section details the implementation of this project, explaining its main software components and techniques used to tackle the development problems. This is an open source project that is hosted in a public GitHub repository\footnote{https://github.com/andrelago13/jarvis}.

\begin{figure}
    \begin{center}
        \includegraphics[width=0.5\textwidth]{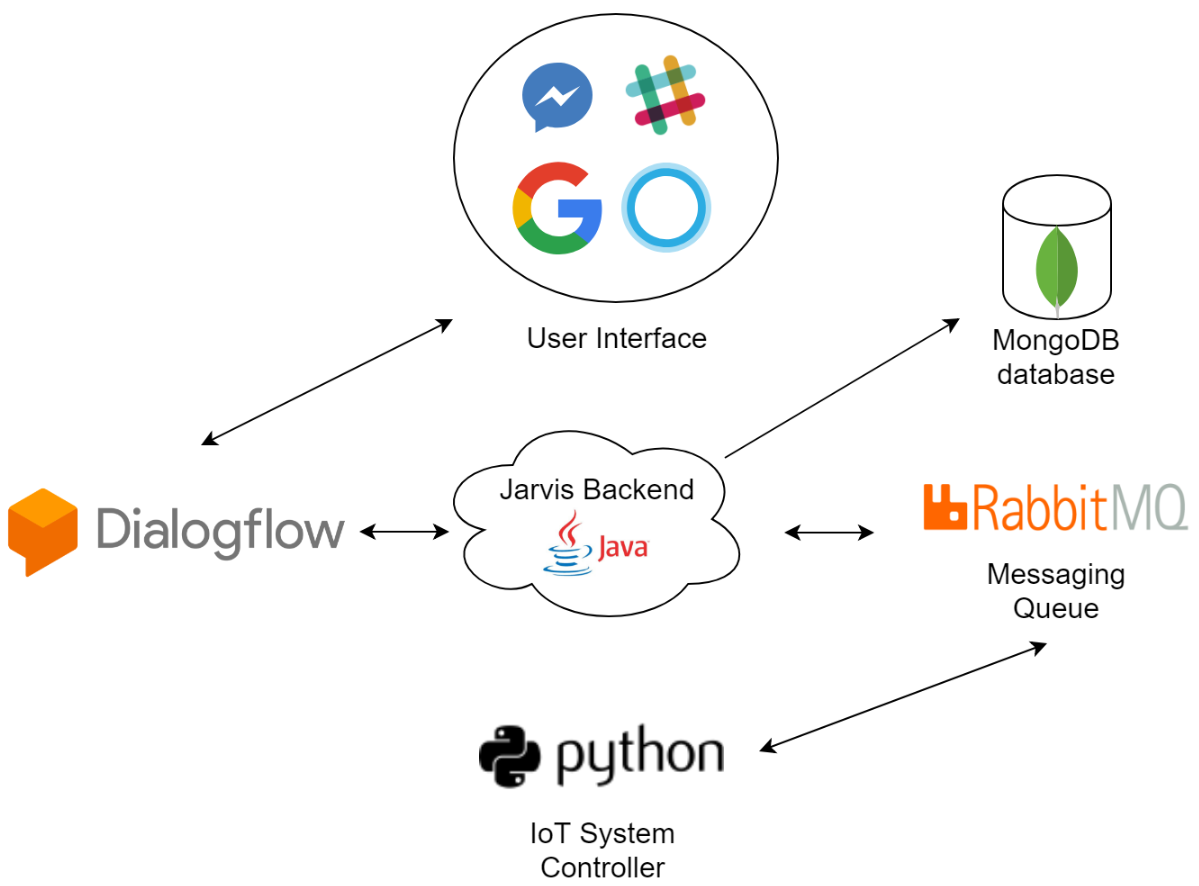}
        \caption{Jarvis overall architectural components.} \label{fig:architecture}
    \end{center}
\end{figure}

Figure~\ref{fig:architecture} presents the high level software components of Jarvis. Each of the components is explained in the following subsections.

\subsection{User Interface}

The user interface is the means through which the user interacts with Jarvis. Due to the use of DialogFlow, there is no need to create specific interfaces for Jarvis as this platform provides built-in integration with multiple popular interfaces. In this case, the main interfaces used were Slack and the Google Assistant so that both text and voice interfaces were covered.

Slack\footnote{\url{https://slack.com/}} is a very powerful team-communication tool used worldwide for team collaboration and efficient communication. Slack provides direct communication with individuals or groups of individuals, but also channeled conversations. In this case, users may create public or private messaging channels in which they can write themed content for other members of the channels to read.

The integration with the Google Assistant is also made easily due to Dialogflow's built-in integration system. With it, the user may use any Google Assistant enabled device with his account to communicate with the bot, only needing to start with a query such as \textit{"Hey Google, talk to Jarvis"}.

Regardless of which of these interfaces is used, the strings representing the exact user query are sent from the interface to DialogFlow’s backend. Because Dialogflow is being used, there is no need to implement any Speech Recognition techniques because they are already implemented. This means that Dialogflow's backend receives raw strings that represent the user queries, which are then analyzed using Natural Language Processing techniques.

\subsection{Dialogflow Backend}

This system receives the user queries from the different user interfaces and parses them into machine understandable code.

Upon receiving a request, there are two things DialogFlow can do: either respond with an automatic response or send the parsed request to a fulfillment backend (in this case, the Jarvis Backend) which will then process the request and return the desired response.

There are a few key concepts that are important to understand in Dialogflow in order to make it operate the best possible way. All the information below can be found in the official Dialogflow documentation\footnote{\url{https://dialogflow.com/docs/getting-started/basics}}.

\begin{easylist}[itemize]
  & \textbf{Fulfillment Backend:} a secondary server that provides a REST API for Dialogflow to send messages to in order to receive the response to the user. This backend receives the result of the processing done by Dialogflow, being able to perform other actions that Dialogflow can't perform (e.g. interact with IoT devices).
  & \textbf{Request:} plain string that represents a user query. An example would be “turn on the living room light”.
  & \textbf{Entity:} symbol that can be represented by different literal strings. As an example, there may be an entity called “toggleable-device” which may be represented by “living room light” or “kitchen light”. Additionally, entities may be represented by other entities, which means that an entity “device” could be represented by the entity “toggleable-device” which then may be represented by certain strings. Entities are represented in DialogFlow with the use of the “@” symbol (“@device”).
  & \textbf{System entity:} a system entity is a regular entity that instead of being manually defined by the user is already defined by the Dialogflow system.
  & \textbf{Intent:} represents a set of example requests that may contain multiple entities. Intents are defined manually and, ideally, each represents one type of query by the user. As an example, an intent named \textit{“Turn on/off device”} may be represented by the requests \textit{“turn the @device on”} and \textit{“turn the @device off”}. In this case, if the request is \textit{“turn the kitchen light on”}, the DialogFlow engine will understand that \textit{“@device”} is represented by \textit{“kitchen light”} and provide that information to the fulfillment backend.
  & \textbf{Context\footnote{\url{https://dialogflow.com/docs/contexts}}:} contexts are used to allow intents or requests to depend on previous instances, enabling for the creation of context-aware interactions. A context is defined by a simple string (e.g. “device-choice”) and can be used as input and/or output for an intent. If an intent has a context as output, the following requests will carry that context in their parameters. On the other hand, if an intent has a context as input, a request will only be parsed into that intent if it carries that context in its parameters (e.g. if a previous intent had that same context as output). Contexts may also carry parameters, which are a simple list of key-value pairs. Whether or not a context is the output of an intent can either be pre-defined in DialogFlow or defined by the fulfillment backend.
\end{easylist}

\begin{figure}
    \begin{center}
        \includegraphics[width=0.8\textwidth]{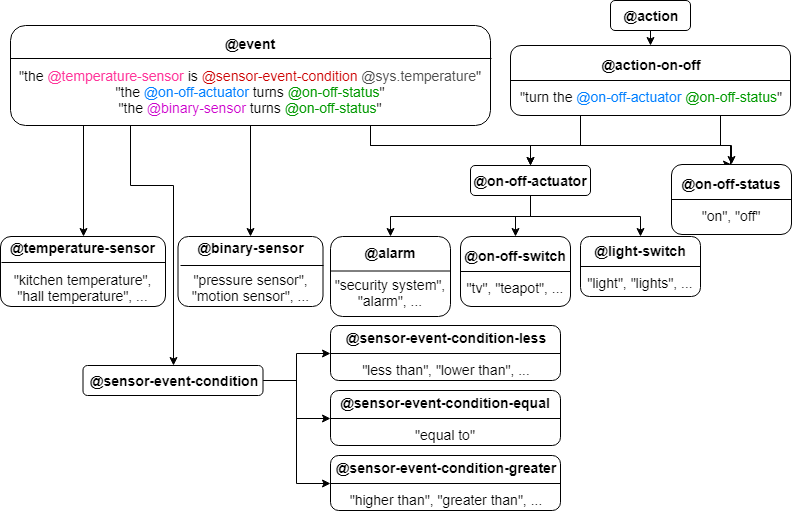}
        \caption{Main entities defined in Jarvis' Dialogflow project.} \label{fig:entities}
    \end{center}
\end{figure}

In the case of Jarvis, there are multiple intents, entities and contexts defined to make the entire system work. Figure~\ref{fig:entities} illustrates the main entities defined for Jarvis in Dialogflow.

The many different intents defined for Jarvis make it possible to support all of its features. The complete list of intents is as follows: Direct Action Intent; Delayed Action Intent; Confirm Thing Choice Intent; Repeating Intent; Event Intent; "Why did something happen?" Intent; Rules Defined Intent; Rules Defined Change Single Rule Intent; Cancel Command Intent; Confirm Cancel Intent. As an example, the Event Intent is explained below:

\begin{easylist}[itemize]  
  & \textbf{Event Intent}
  && \textbf{Usage:} creates an action that is performed upon a certain event, such as an activity of another device or a change of a device's status.
  && \textbf{Definition:}
  &&& "@action:action when @event:event"
  && \textbf{Example:}
  &&& \textit{"Turn on the bedroom light when the living room light turns off"}
\end{easylist}

Having the intents defined, the Dialogflow backend takes requests and builds JSON objects that contain all the information related to the intent parsed as well as the entities that are present in it. These objects are then sent to the Jarvis backend for further processing.

\subsection{Jarvis Backend}

The Jarvis backend is a JavaEE\footnote{\url{http://www.oracle.com/technetwork/java/javaee/overview/index.html}} application that provides a REST API to which Dialogflow requests are sent via a POST URL to compute the replies to user queries as well as perform the required actions on IoT devices.

For each of the intents mentioned in the previous section, the backend has an equivalent class that is responsible for parsing the request and generating a response. Upon receiving a request, each of the intent classes is responsible for validating the request parameters to make sure that not only all required parameters are present (e.g. device names or desired action) but also that the specified devices, if any, are clear and unique. If the request has errors, an appropriate and explanatory response should be returned. If the parameters are valid but the intended device is unclear (e.g. user wants to turn on the "light" but there is a "living room light" and a "bedroom light"), the device specification context should be added and the response should ask the user to specify the desired device.

\begin{figure}
    \begin{center}
        \includegraphics[width=0.6\textwidth]{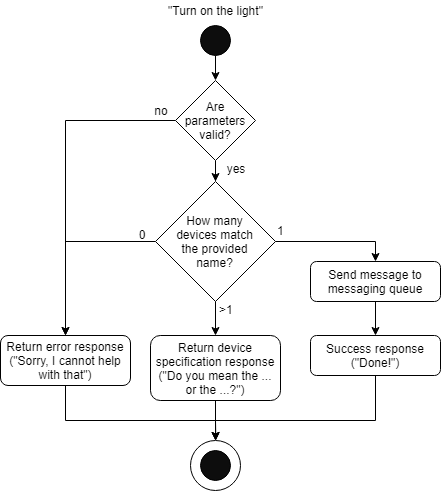}
        \caption{Activity diagram for the parsing of the query "\textit{turn on the light}".} \label{fig:intent-activity}
    \end{center}
\end{figure}

To perform the logs of commands and user commands, all actions that can be performed are represented as commands, therefore using the Command\footnote{\url{http://w3sdesign.com/}} design pattern. Each command has "execute" and "undo" methods, implemented in a way that calling "undo" after "execute" should return the system to its original state.

With most intents, such as direct actions or "why did something happen?" queries, the effects are immediate and require no delayed activity from the backend, except for period actions, rules and events. These require a special engineering so that they can perform actions on the backend without the need of a request to trigger them.

A period action is an action that must be done and then undone after a certain period of time. This can happen as a result of a query such as "\textit{turn on the light from 4pm to 5pm}". As a result of this sort of query, a command is generated. This command includes an implementation of a state machine in which the state tells whether the first action (4pm) was not executed yet, the first action was executed but the second (5pm) wasn't or both have been executed. Therefore, when the command is executed, it schedules a Java thread to be run at 4pm, at which time the command is executed again, changing its state and executing the 4pm action. At this point, a second thread is scheduled for 5pm to execute the second action and move the command to its final state. The use of the Command pattern is useful as it abstracts from the engine the notion of direct or delayed commands.

The same concept is used for system rules, which are used for queries such as "\textit{turn on the light everyday at 5pm}". In this case, at first, a thread is scheduled for the following "5pm" to execute the command for the first time. From then on, everytime the command is executed it schedules a new thread for the following 5pm, but never leaves this state of rescheduling. The only way to cancel this action is to call the "undo" method which cancels the currently scheduled thread.

This thread scheduling mechanism does not work for events because the time at which these are executed is not known. Events are a result of queries such as "\textit{turn on the light when the sensor is activated}". In this case, it is necessary for the system to listen to the messaging queues of the device actions and states (these queues are explained below in another section). As will be explained below, each device has messaging queues that relate to its actions and state changes. Messages are published in the latter by the device controller everytime its state is changed, which means that listening to those queues allows a listener to be notified of state changes on the device.

On the backend, it wouldn't make sense to create a new listener for every event that uses a certain condition as that would put more load on the messaging queue layer of the backend. Instead, a different mechanism is used. Upon startup of the backend, a listener is created for each event queue of each device. These are simple classes that are responsible for receiving messages published in those queues and notifying the Jarvis backend engine that those messages were received.

The engine then uses the Observer\footnote{\url{http://w3sdesign.com/}} design pattern to allow commands to add observers to these messages. For example, the query "\textit{turn on the light when the sensor is activated}" would add an observer that would look for messages on the sensor's event queue with the value "on". Because of that, the engine would notify this observer of all messages on that queue leading to the event being executed. The advantage of this technique is that it allows the engine to be a centralized point through which all messages go through and with which all observers must be registered.

Other patterns are used across the solution such as Builder (for creating IoT device classes), Singleton (for database and messaging queue networked connections) and Adapter (for abstracting the networked interactions with the messaging queue for the IoT devices).

One of the main features of this project are the causality queries, the requests through which the user asks why something happened (e.g. "\textit{why did the light turn on?}". To implement them, the command classes also have a method that determines whether or not they could have caused a certain condition to be true. Because of that, a simple solution to this problem is to, given the queried condition, look through the log of commands and user commands to understand what could have caused that condition. However, this approach does not work for all the use cases, for example, when, at the moment of the query, multiple rules may have caused the condition to be true. In these cases, it is not enough to return the latest logged command that could cause the queried condition. Instead, there are two possible approaches:

\begin{easylist}[itemize]
  & \textbf{Return earliest possible cause:} if multiple rules/events could have caused the queried condition to be true, choose the one that happened first as that was the cause for the condition to be true in the first place.
  & \textbf{Use an heuristic to calculate most relevant cause:} instead of assuming the best answer is the earliest cause, use an heuristic to evaluate all the possible causes in terms of relevance and choose the most relevant. This calculation can use the nature of the cause (caused by user vs. caused by rule), order by which causes happened, among others.
\end{easylist}

In the context of this project the implemented solution was to use the earliest possible cause, but it is important to mention that the alternative would be just as valid and possibly more useful to the user.

It is also possible that a chain of interconnected rules caused the queried condition to be true. In this case, it is possible to either reply with the complete chain of events, the latest possible cause or engage in a conversation through which the user can explore the full chain of events chronologically (e.g. by saying "\textit{tell me more}" to explore the chain).

In the context of this project, the solution uses the approach of returning the latest possible cause. However, with the mindset of creating a helpful and fully capable bot, the best approach would be to engage in a conversation with the user, not only because it would allow the user to obtain more information at his own pace but also because it would be possible to make changes to the rules as he goes. For example, if the user is having that conversation and he decides to change one of the rules, he can just interrupt the conversation to modify the rule. With the current approach, that is only possible if the answer contains a single rule or cause to the condition which in reality is not always the case.

\subsection{Database, Messaging Queue and IoT System Controllers}

These components are not crucial to the functioning of Jarvis so they will be described very briefly in this Section. These fall slightly out of the scope of the project, however they are useful to understand how the entire system works.

The chosen database technology was MongoDB\footnote{\url{https://www.mongodb.com/}} due to the fact that it is a popular document store database. This database was used to store the list of available devices as well as a log of the commands executed by the Jarvis backend, which was useful for answering causality queries.

The chosen messaging service was RabbitMQ\footnote{\url{https://www.rabbitmq.com/}} due to the fact that it facilitates a Publisher-Consumer architecture while providing a scalable and reliable solution. Different messaging queues were created on the service with two purposes. On one hand, action queues were used for the Jarvis backend to publish messages to create actions on the IoT devices (e.g. publishing "on" to turn on a light). On the other hand, event queues were used for the devices to publish state changes (e.g. a new temperature detected in a sensor), which in turn were used for events in the Jarvis engine.

The IoT system controllers are the entity that falls further away from the scope of the project. These were simple Python scripts that were used to read and write messages from the messaging queues to appropriately act on the IoT devices using Python's hardware libraries (in this project's case, the RPi.GPIO\footnote{\url{https://pypi.org/project/RPi.GPIO/}} since RaspberryPi devices were used).

\section{Validation}

To evaluate the success of this project, two measures were taken to understand how it may outperform other systems due to the features it provides as well as how easy and intuitive it is to use. For that, a list of simulated scenarios was elaborated to understand how the implemented features are present in currently available alternatives for smart space management. Then, a user study was done to evaluate the project's ease of use and what users may think of it.

\subsection{Simulated Scenarios}

Table~\ref{table:comparison} shows how the developed features compare to current alternatives of visual platforms and conversational assistants, in this case, respectively, Node-RED and the Google Assistant (these were chosen as examples as they are among the most complete alternatives in each field).

It is important to clarify that "One-time action w/unclear device" refers to actions like "\textit{turn on the light}" with which Jarvis asks the user to clarify which device he means based on the given name ("\textit{do you mean the bedroom or living room light?}"). "Cancel last command" refers to that actual command being said, which should cause the last command to be undone, be it an action or a rule creation. Finally, "Rules defined for device" refers to the user querying Jarvis what the defined rules for a device are (e.g. "\textit{what rules are defined for the bedroom light?}").

\begin{table}
    \centering
    \begin{tabular}{ | c | c | c | c |}
    \hline
    Scenario & Jarvis & Google Assistant & Node-RED \\ \hline

    One-time action &
    \includegraphics[width=20pt]{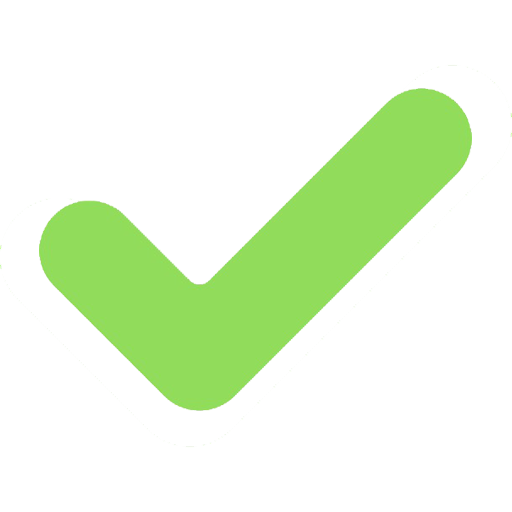} &
    \includegraphics[width=20pt]{checkmark} &
    \includegraphics[width=20pt]{checkmark} \\ \hline
    
    One-time action w/unclear device &
    \includegraphics[width=20pt]{checkmark} &
    \includegraphics[width=20pt]{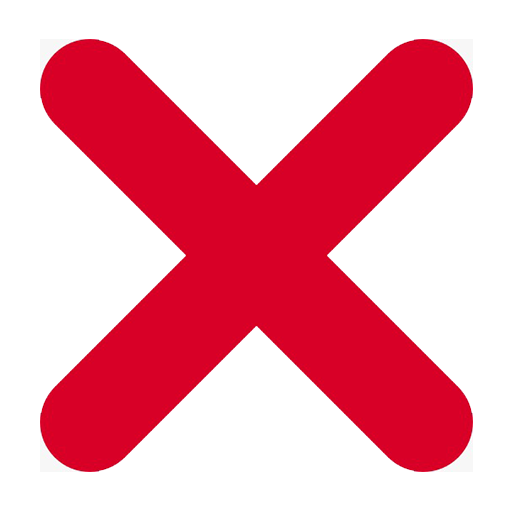} &
    \includegraphics[width=20pt]{cross} \\ \hline
    
    Delayed action &
    \includegraphics[width=20pt]{checkmark} & 
    \includegraphics[width=20pt]{cross} &
    \includegraphics[width=20pt]{checkmark} \\ \hline
    
    Period action &
    \includegraphics[width=20pt]{checkmark} & 
    \includegraphics[width=20pt]{cross} &
    \includegraphics[width=20pt]{checkmark} \\ \hline
    
    Daily repeating action &
    \includegraphics[width=20pt]{checkmark} & 
    \includegraphics[width=20pt]{cross} &
    \includegraphics[width=20pt]{checkmark} \\ \hline
    
    Daily repeating period action &
    \includegraphics[width=20pt]{checkmark} & 
    \includegraphics[width=20pt]{cross} &
    \includegraphics[width=20pt]{checkmark} \\ \hline
    
    Cancel last command &
    \includegraphics[width=20pt]{checkmark} & 
    \includegraphics[width=20pt]{cross} &
    \includegraphics[width=20pt]{cross} \\ \hline
    
    Event rule &
    \includegraphics[width=20pt]{checkmark} & 
    \includegraphics[width=20pt]{cross} &
    \includegraphics[width=20pt]{cross} \\ \hline
    
    Rules defined for device &
    \includegraphics[width=20pt]{checkmark} & 
    \includegraphics[width=20pt]{cross} &
    \includegraphics[width=20pt]{cross} \\ \hline
    
    Causality query &
    \includegraphics[width=20pt]{checkmark} & 
    \includegraphics[width=20pt]{cross} &
    \includegraphics[width=20pt]{cross} \\ \hline
    \end{tabular}

    \caption{Simulated scenarios fulfillment in Jarvis, Google Assistant and Node-RED}
    \label{table:comparison}
\end{table}

The important conclusion to withdraw from the table is that Jarvis provides a lot of features that aren't present in either the Google Assistant or Node-RED. Obviously both of these products do a lot more than these features, but especially with the Assistant the advantage is clear since the only kind of smart space feature it supports is the one shown in the table.

Overall, these studies prove that it is possible to bring some of the features available in Node-RED to a conversational interface, as well as bring even more features that provide increased complexity and management.

\subsection{User Study}\label{sec:userstudy}

The study was done with 17 participants with ages ranging from 18 to 51. The main goal on obtaining the test participants was to obtain a significant age range but also to include mostly people without a background in technology to understand whether the language capabilities of Jarvis are enough for the users to interact with it. Because of that, 14 of the participants did not have a background in software development although they were familiar with different technologies such as smartphones and the internet.

Each participant was given a set of 5 tasks (1 control task and 4 study tasks) that they were supposed to get done with the help of Jarvis, using the Google Assistant as the user interface. The only instructions given were that they must talk to the phone in the way that feels the most natural to them in order to complete the tasks at hand. There were 2 sets of tasks that participants were assigned to randomly.

Besides the set of tasks, participants were given the list of IoT devices available in the simulated smart house they would be attempting to manage through Jarvis. To increase the diversity and reduce the bias of the study, two different sets of devices and tasks were created, so that different smart house topologies were tested. The participants were assigned one of the test sets randomly.

For each of the tasks, the study administrator would take note of whether each participant was able to complete each of the tasks, the time it took to complete the task and the count of unsuccessful queries. This count was made separately for queries that were not understood by the Assistant's speech recognition capabilities, for queries the user mispronounced and for queries that were correct but Jarvis was unable to process. After completing the tasks, the administrator would explain to participants that an alternative to Jarvis to perform the tasks at hand was a non-conversational visual interface such as Node-RED, giving an example of how the same tasks would be performed using that tool. After that, the administrator would ask participants what they believe are the advantages of Jarvis over such a tool, if they find any, and whether they would prefer Jarvis over any non-conversational tool. Finally, the participants would be asked what they think could be improved about Jarvis and the way it handles smart space management.

Table~\ref{table:studyresults} compiles the results observed during the study. Each row represents one of the tasks given to participants. The meaning of each column is as follows:

\begin{easylist}[itemize]
    & \textbf{"Task:"} number of the task (0-4) and the task set number in parenthesis (1/2).
    & \textbf{"Done:"} percentage of participants that were able to complete the task successfully.
    & \textbf{"Time:"} average ("\textit{Avg}") and median ("\textit{Med}") time, in seconds, that participants took to complete the task.
    & \textbf{"IQ (Ast):"} average ("\textit{Avg}") and median ("\textit{Med}") times that queries were incorrect due to the Google Assistant not properly recognizing the user's speech (due to microphone malfunction, background noise, ...).
    & \textbf{"IQ (User):"} average ("\textit{Avg}") and median ("\textit{Med}") times that queries were incorrect due to the user not speaking a valid query (e.g. saying "\textit{Turn up the lighting}").
    & \textbf{"IQ (Jvs):"} average ("\textit{Avg}") and median ("\textit{Med}") times that queries were incorrect due to Jarvis not recognizing a valid query.
    & \textbf{"IQ:"} average ("\textit{Avg}") and median ("\textit{Med}") times queries were incorrect (sum of "\textit{IQ (Ast)}", "\textit{IQ (User)}" and "\textit{IQ (Jvs)}").
\end{easylist}

\begin{table}
    \centering
    \begin{tabular}{ | c | c | c | c | c | c | c | c | c | c | c | c |}
    \hline
    \multicolumn{2}{|c|}{} & \multicolumn{2}{|c|}{Time} & \multicolumn{2}{|c|}{IQ (Ast)} & \multicolumn{2}{|c|}{IQ (User)} & \multicolumn{2}{|c|}{IQ (Jvs)} & \multicolumn{2}{|c|}{IQ} \\ \hline
    Task & Done & Avg & Med & Avg & Med & Avg & Med & Avg & Med & Avg & Med \\ \hline
    
    0 (1) & 94\% & 6.4 & 6 & 0.13 & 0 & 0.25 & 0 & 0.13 & 0 & 0.5 & 0 \\ \hline
    1 (1) & 94\% & 7.1 & 7 & 0.38 & 0 & 0.5 & 0.5 & 0 & 0 & 0.5 & 0.5 \\ \hline
    2 (1) & 88\% & 10 & 10 & 0.75 & 0.5 & 0.63 & 0.5 & 0.25 & 0 & 1 & 1 \\ \hline
    3 (1) & 100\% & 20 & 19.5 & 0.13 & 0 & 0.13 & 0 & 0.75 & 1 & 1 & 1 \\ \hline
    4 (1) & 94\% & 9 & 8 & 0.25 & 0 & 0.38 & 0 & 0 & 0 & 0.63 & 0 \\ \hline
    0 (2) & 100\% & 6.4 & 6 & 0.33 & 0 & 0 & 0 & 0.33 & 0 & 0.67 & 0 \\ \hline
    1 (2) & 94\% & 7.6 & 7 & 0.11 & 0 & 0 & 0 & 0.44 & 0 & 0.56 & 0 \\ \hline
    2 (2) & 100\% & 9.9 & 10 & 0 & 0 & 0.11 & 0 & 0.78 & 1 & 0.89 & 1 \\ \hline
    3 (2) & 88\% & 19.44 & 19 & 0.33 & 0 & 0.33 & 0 & 0.22 & 0 & 0.89 & 1 \\ \hline
    4 (2) & 100\% & 8.33 & 8 & 0.33 & 0 & 0.22 & 0 & 0.22 & 0 & 0.78 & 1 \\ \hline
    \end{tabular}

    \caption{User Study results (task completion rate, task time and incorrect queries).}
    \label{table:studyresults}
\end{table}

Based on the results presented in Table~\ref{table:studyresults}, an \textbf{analysis} of what they represent and how they may be interpreted is made below.

The complexity of the queries increases from task 0 to task 3 since the queries require more words or interactions. This is reflected by the task time that also increases from task 0 to task 3 in both the average and median values and in both task sets. Other than that, it is hard to use the task times as means of comparison to visual tools since no comparison study was done. However, since using a visual tool such as Node-RED involves moving to a computer, opening the platform, dragging blocks and then configuring their options, we consider that in any case the overall task would take longer in a visual tool than it does with Jarvis.

%The numbers related to incorrect queries also bring some interesting conclusions about the system's performance. The table shows that there were some incorrect queries for the Assistant which means that the Assistant's speech recognition failed to correctly translate what the participants said. However, this was mostly due to the participant's pronounciation since none of the speakers had English as their main language. This does not have implications to the evaluation of Jarvis, but it does indicate that this sort of systems might be harder to use if they do not support the user's native language.

It can also be noted that there were a few instances of incorrect queries that were the user's fault ("\textit{IQ (User)}"). These were cases where user queries were gramatically incorrect and therefore did not match the sample queries defined in Dialogflow. For example, a query like "\textit{Turn on lights}" is not a considered the user's fault since the correct one would be "\textit{Turn on the lights}", however, it still carries enough information to understand what the user's intent is. More reflexion on these cases will be done below.

There is a significant number of incorrect queries due to Jarvis ("\textit{IQ (Jvs)}"), meaning that the user's query was valid but its meaning was not understood by Jarvis. This can be caused by either a mispronounciation of a device's name (e.g. saying "\textit{Turn on the living room light\textbf{s}}" instead of "\textit{Turn on the living room light}") or a sentence structure that is valid but is not recognized due to the sample queries inserted into Dialogflow. This possibly represents the most serious downside of this project, but its severity and possible solutions are discussed below.

It is also important to note that despite the numbers for incorrect queries, the success rate of all tasks is very high, which indicates that the system is intuitive enough to be used without previous instruction or formation. This proves that not only the features work on the smart space management side, but also that the conversational interface is well designed and works as expected.

All of the reflexions made above are reflected by the participants' answers to the questions made, since they reiterate that Jarvis is easy to use and provides "accessibility" and "commodity". This is very important since it is the deciding factor for participants to prefer Jarvis over a non-conversational interface, which they still consider to have some advantages over Jarvis.

The analysis of the study results made above is very important to analyze the possible \textbf{validation threats} that they bring upon this project:

\begin{easylist}[itemize]
    & \textbf{Not all incorrect queries should fail:} a query like "\textit{Enable the lights}" might not be very common or correct, but it still carries enough information so that a human would understand what its goal is. Including gramatically incorrect queries in Dialogflow's system would make Jarvis accept these partially incorrect queries, making it more flexible to these scenarios and therefore improve the user experience.

    & \textbf{Poor flexibility towards correct queries:} if a device's name is "bedroom light", if the user says "\textit{turn on the bedroom light\textbf{s}}" the query would faild due to the plural being used. This may cause a bad user experience in cases of slight mispronounciation and can easily be fixed by both providing slightly incorrect queries to the Dialogflow sample requests or by improving the device search in the Jarvis backend to not only look for exact device name matches.

    & \textbf{Reduced diversity of valid queries:} the study showed that there are many different ways to say valid and supported queries that do not work with Jarvis due to them not being covered by the Dialogflow samples. Again, this creates a poor user experience and could easily be improved by widening the range of sample requests in Dialogflow.
\end{easylist}

Overall, the high success rate of the tasks showed that the system is intuitive enough to be used by people that do not posess high technological knowledge, however, it also showed that improvements can be made in terms of making the system more flexible to the many different ways, correct or incorrect, that users can find to give the same commands.

\section{Conclusions and Future Work}

This final section of the article presents the main contributions and conclusions of this project, ending with a description of the future work that is planned.

\subsection{Main Contributions}

The main contributions of Jarvis are focused on new IoT management features for conversational assistants. Because of that, the main contributions of this project are the following:

\begin{easylist}[itemize]
    & \textbf{Delayed, period and repeating actions:} the ability for users to perform queries such as "\textit{Turn on the light in 5 minutes}" or "\textit{Turn on the light everyday at 8am}".
    
    & \textbf{Use contextual awareness for more natural conversations:} the ability to have conversations that last for multiple sentences to provide a more intuitive converstation. This is started by queries such as "\textit{What rules do I have defined for the living room light?}".
    
    & \textbf{Event management:} this is used to create interactions and dependencies between devices that might not necessarily know that each other exists. This is used for queries such as "\textit{Turn on the light when the motion sensor is activated}".
    
    & \textbf{Causality queries:} the ability to understand how the system operates simply by asking it, with queries such as "\textit{Why did the light turn on?}".
\end{easylist}

\subsection{Conclusions}

The main conclusion of this project is that it is possible to use a conversational interface to provide a useful management tool for smart spaces. As was mentioned in Section~\ref{sec:introduction}, current conversational bots already have the capacity to interact with IoT devices, so an integration of the features of Jarvis would make them more complete and useful to users.

At the same time, through stages of the development such as the user study mentioned in Section~\ref{sec:userstudy} it becomes clear that natural language is a tricky part of the process. The user study showed how hard it is to predict the many user queries that have the same intrinsic meaning, and this problem would only be increased if multiple languages were targeted. This wouldn't have implications on the complexity of tasks that the bot does, but considering multiple languages would bring additional problems not only for defining user queries but also for establishing the representation of entities such as devices.

Another conclusion drawn from the use of a conversational interface is that there are additional problems when the responses to the user are two long. Comments on the user study state that when the responses provided by the system are too long they may get harder to understand since there is more information to be conveyed. A possible solution for this problem would be to use a hybrid interface that provides both visual and audio interactions, but there could be other approaches such as an interactive dialogue that shortens sentences.

We believe that the contribution of Jarvis is clear, as it outperforms current conversational assistants in terms of supported features while simultaneously being easier to use for users (as seen by the user study).

\subsection{Future Work}\label{sec:futurework}

An interesting future feature for Jarvis would be alias commands. These are shortcuts to existing commands, so that, for example, saying "\textit{heat the room}" would be the same as saying "\textit{set the living room temperature to 20 degrees}". This can be a very useful feature as it enables users to shorten commands and make the system more intuitive for them.

Another interesting exploration that could be made would be the study of undocumented dependencies in chains of events. Currently, if there is a chain of events that is triggered and where the action of an event is the direct cause for the following event, asking Jarvis why the final action happened returns the whole chain of events (e.g. if the bedroom light turns on when the living room light turns on and the latter turns on at 9 a.m., then asking why the bedroom light turned on at 9:05 a.m. would correctly return that it is because it's 9 a.m.). An undocumented dependency would happen if, for example, one event turns off the light when another event is triggered by it getting dark. Making Jarvis able to understand the correlation of these 2 events even if their effects are not directly represented by system rules would make it more accurate and useful in answering causality queries.

% Section~\ref{sec:usecase15} mentions causality queries for undocumented relations of events. This was not possible in the span of the project mostly because it involves serious considerations on how the induction of relationships between events can be done. One possible approach to this would be considering events that happened close to each other to possibly be related, but this could not be a sufficient criteria in many scenarios. Therefore, we believe that a further study of event relations that are not explicit to be an important future step for Jarvis and for conversational smart space management tools.

The addition or removal of devices was not considered for the development of this project. However, it would obviously be important for a user product in order to allow users to dinamically add or remove devices to their smart spaces. This would be very useful for the evolution of Jarvis and another big step towards making conversational systems able to fully manage smart spaces.

As was discussed in Section~\ref{sec:userstudy}, the user study that was made revealed that the system's overall flexibility could be improved by increasing the sample requests in Dialogflow since many of the failed queries in the study carried enough information for them to be understood. Therefore, improving the selection of sample requests to include partially invalid queries or rephrasing the current ones can improve the system's perfomance by reducing its error rate and therefore improving the user experience.

A useful future feature for Jarvis would be a "\textit{What can you do?}" query that prompts the system to explain to a user what it can do. This can be used to create an interactive dialogue that can be used as a tutorial for a beginner user which in turn makes the system even easier to use. We believe this would be a very important feature for users that are less acquainted with technology, since they learn the types of queries and supported features more intuitively.

Finally, we believe a major feature that could be made available in conversational assistants would be complex boolean logic. For example, when defining event rules, it would be useful to use multiple conditions with the "and" or "or" boolean operators. An example of this feature would be the query "\textit{Turn on the bedroom light if the motion sensor is activated and it's after 9pm}", where both conditions would have to be true in order to the action to be executed. This would provide conversational assistants with increased complexity and usefulness, which in turn would make them even more powerful tools for users.

Overall, considering Jarvis' goal of making IoT easier to use for users that are not very comfortable with technology, many of the future work tasks mentioned in this section aim to make interactions even more intuitive and easier to use, without requiring a technical knowledge of how the devices operate.

\nocite{Rasch2014, VanderMeulen2017, bureaustatistics, Suresh2014, Chen2014, randall2003living, maruti, olgadavydova, joebirch}

\bibliographystyle{splncs04}
\bibliography{mybibliography}

\begin{thebibliography}{10}
\providecommand{\url}[1]{\texttt{#1}}
\providecommand{\urlprefix}{URL }
\providecommand{\doi}[1]{https://doi.org/#1}

\bibitem{bureaustatistics}
{Average hours per day spent in selected activities by sex and day}. Bureau of
  Labor Statistics, retrieved from
  \url{https://www.bls.gov/charts/american-time-use/activity-by-sex.htm} (2017)

\bibitem{joebirch}
Birch, J.: {Exploring Dialogflow: Understanding Agent Interaction}. Medium,
  retrieved from
  \url{https://medium.com/@hitherejoe/exploring-dialogflow-understanding-agent-interaction-8f3323e3b738}
  (October 2017)

\bibitem{Chen2014}
Chen, S., Xu, H., Liu, D., Hu, B., Wang, H.: {A vision of IoT: Applications,
  challenges, and opportunities with China Perspective} (Aug 2014).
  \doi{10.1109/JIOT.2014.2337336}

\bibitem{olgadavydova}
Davydova, O.: {25 Chatbot Platforms: A Comparative Table}. Medium, retrieved
  from
  \url{https://chatbotsjournal.com/25-chatbot-platforms-a-comparative-table-aeefc932eaff}
  (May 2017)

\bibitem{VanderMeulen2017}
van~der Meulen, R.: {Gartner Press Release, Gartner Says 8.4 Billion Connected
  "Things" Will Be in Use in 2017, Up 31 Percent From 2016} (2017).
  \doi{10.1017/CBO9781107415324.004}

\bibitem{Rahul2017}
Mourya, R.: {IoT applications spanning across industries}. IBM Blog, available
  at
  \url{https://www.ibm.com/blogs/internet-of-things/iot-applications-industries/}
  (Apr 2017)

\bibitem{randall2003living}
Randall, D.: Living inside a smart home: A case study. In: Inside the smart
  home, pp. 227--246. Springer (2003)

\bibitem{Rasch2014}
Rasch, K.: {An unsupervised recommender system for smart homes}. Journal of
  Ambient Intelligence and Smart Environments  \textbf{6},  21--37 (2014).
  \doi{10.3233/AIS-130242}

\bibitem{Suresh2014}
Suresh, P., Daniel, J.V., Parthasarathy, V., Aswathy, R.H.: {A state of the art
  review on the Internet of Things (IoT) history, technology and fields of
  deployment}. In: 2014 International Conference on Science Engineering and
  Management Research (ICSEMR). IEEE (Nov 2014).
  \doi{10.1109/ICSEMR.2014.7043637}

\bibitem{maruti}
Techlabs, M.: {Exploring Dialogflow: Understanding Agent Interaction}. Medium,
  retrieved from
  \url{https://chatbotslife.com/which-are-the-best-on-site-chatbot-frameworks-3dbf5157fb57}
  (April 2017)

\bibitem{Xia2012}
Xia, F., Yang, L.T., Wang, L., Vinel, A.: {Internet of Things}. INTERNATIONAL
  JOURNAL OF COMMUNICATION SYSTEMS Int. J. Commun. Syst. Int. J. Commun. Syst
  \textbf{25}(25) (2012). \doi{10.1002/dac}

\end{thebibliography}




% LNCS Homepage, \url{http://www.springer.com/lncs}. Last accessed 4

%\end{thebibliography}
\end{document}